\documentclass[aps,prl,twocolumn,showpacs,amssymb]{revtex4}
\usepackage{graphicx}

\usepackage[latin1]{inputenc}
\usepackage[rightcaption]{sidecap}
\usepackage{amssymb}
\usepackage{amsmath}
\usepackage{color}
\usepackage{subfigure}
\usepackage{floatflt}
\usepackage{units}
\usepackage{lscape}
\begin{document}
%#####################################################################
\title{Femtosecond spin current pulses generated by the non-thermal spin-dependent \\ Seebeck effect and interacting with ferromagnets in spin valves}

\author{Alexandr~Alekhin$^{1}$}\altaffiliation[Present address: ] {IMMM UMR CNRS 6283, Université du Maine, Avenue Messiaen, 72085 Le Mans, France.}
\author{Ilya~Razdolski$^{1}$}
\author{Nikita~Ilin$^{1}$}\altaffiliation[Present address: ] {Moscow State Technical University of Radioengineering, Electronics and Automation, Prospect Vernadskogo 78, 119454 Moscow, Russian Federation.}
\author{Jan~P.~Meyburg$^{2}$}
\author{Detlef~Diesing$^{2}$}
\author{Vladimir~Roddatis$^{3}$}
\author{Ivan~Rungger$^{4}$} \altaffiliation[Present address: ] {National Physical Laboratory, Teddington, TW11 0LW, United Kingdom.}
\author{Maria~Stamenova$^{4}$}
\author{Stefano~Sanvito$^{4}$}
\author{Uwe~Bovensiepen$^{5}$}
\author{Alexey~Melnikov$^{1,6}$} \email[Corresponding author: ] {melnikov@fhi-berlin.mpg.de}

\affiliation{$^{1}$Department of Physical Chemistry, Fritz Haber Institute of the Max Planck Society, Faradayweg 4-6, 14195 Berlin, Germany\\$^{2}$Faculty of Chemistry, University of Duisburg-Essen, Universitätsstr. 5, 45117 Essen, Germany\\$^{3}$Institute of Materials Physics, University of Göttingen, Friedrich-Hund-Platz 1, 37077 Göttingen, Germany\\$^{4}$School of Physics and CRANN, Trinity College Dublin, Dublin 2, Ireland\\$^{5}$Faculty of Physics, University of Duisburg-Essen, Lotharstr.~1, 47057 Duisburg, Germany\\$^{6}$Institute of Physics, Martin Luther University Halle-Wittenberg, Von-Danckelmann-Platz 3, 06120 Halle, Germany}

\date{\today}

\begin{abstract}
Using the sensitivity of magneto-induced second harmonic generation to spin currents (SC), we demonstrate in Fe/Au/Fe/MgO(001) pseudo spin valves the generation of 250 fs-long SC pulses. Their temporal profile indicates that superdiffusive hot electron transport across a sub-100~nm Au layer is close to the ballistic limit and the pulse duration is primarily determined by the thermalization time of laser-excited hot carriers in Fe. Considering the calculated spin-dependent Fe/Au interface transmittance we conclude that a non-thermal spin-dependent Seebeck effect is responsible for the generation of ultrashort SC pulses. We also show that hot electron spins rotate upon interaction with non-collinear magnetization at the Au/Fe interface, which holds high potential for future spintronic devices.
\end{abstract}


\maketitle

%#####################################################################

Expansion of spintronics towards ultrashort timescales requires (i) new techniques for generation, control, and monitoring of ultrashort spin current (SC) pulses and (ii) fundamental understanding of their interaction with magnetic layers. High mobility of electrons makes metals promising materials for prototype devices. It is known that SC can be generated by the temperature gradient in the bulk of a ferromagnetic metal (FM) \cite{Uchida_Nature08} or across the interface to a normal metal (NM) \cite{Slachter_NatPhys10} employing the spin-dependent Seebeck effect, which allows for $\sim100$~ps SC pulses \cite{Choi_Cahill_NatPhys15}. Direct excitation of spin-polarized hot carriers (HC) in FM by ultrashort laser pulses can substantially reduce the SC pulse duration. Spin transfer on femtosecond time scales has been demonstrated by time-resolved non-linear magneto-optics \cite{Melnikov_PRL11,Kampfrath_NatNano13}. The theory of \emph{superdiffusive} HC transport was developed \cite{battiato_PRL10,battiato_PRb12} and used to describe the ultrafast magnetization dynamics in metallic bi- \cite{Stamm_NatMat07,Eschenlohr_NatMat13} and tri-layers \cite{Rudolf_NatCom12,Turgut_PRL13} investigated with XUV and soft x-ray probes.

To obtain insight into the microscopic dynamics of SC pulses is challenging because the available experimental techniques \cite{Stamm_NatMat07,Eschenlohr_NatMat13,Rudolf_NatCom12,Turgut_PRL13,Melnikov_PRL11,Kampfrath_NatNano13} monitor either not the SC pulse but its effect on a material (e.g. transient magnetization) or both in an entangled form. Therefore, the open question whether the HC dynamics in FM, the scattering in NM layer, or the properties of FM/NM interface determine the SC pulse duration and amplitude requires new experimental concepts. The core of such concepts is the direct detection of SC which can be realized by the magneto-induced second harmonic (SH) generation (mSHG) in centrosymmetric media. In addition to the well known interface contributions, mSHG contains bulk terms proportional to the current and SC as demonstrated in GaAs \cite{Ruzicka_PRL12,Werake_NatPhys10} and multilayer graphene \cite{Bykov_PRB12}.

Here we demonstrate ultrashort 250~fs-long SC pulses generated by laser excitation of a Fe/Au/Fe heterostructure employing mSHG detection. We conclude that the SC generation originates from a \emph{non-thermal spin-dependent Seebeck effect} (NSS) based on the spin- and energy-dependent transmittance of the FM/NM interface for non-equilibrium HC. Furthermore, studying the propagation of spin-polarized HC across the heterostructure, we show that HC spins orthogonal to the FM magnetization rotate upon their interaction with the Au/Fe interface. This implies that the transmitted and reflected HC gain opposite spin components, similar to the spatial spin separation known from the Stern-Gerlach experiment.

\begin{figure} \centering
 \includegraphics[width=1\columnwidth]{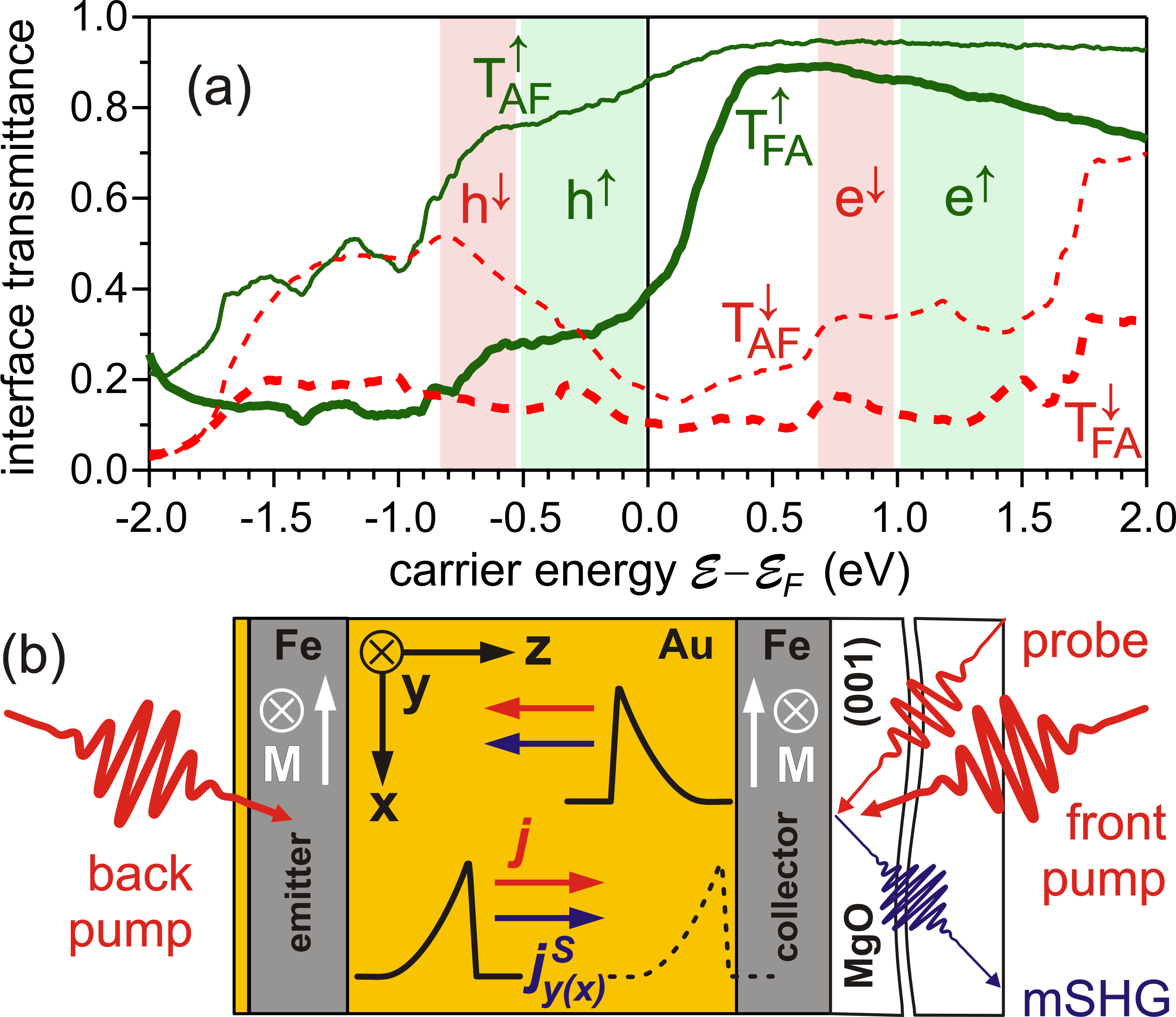}
\caption{\label{Fig1} (a) Calculated momentum-averaged transmittance of the Fe/Au interface $T^\uparrow$ for majority (solid green curves) and $T^\downarrow$ for minority (dashed red) carriers moving from Fe to Au ($T_{FA}$) and from Au to Fe ($T_{AF}$) (thick and thin curves, respectively \cite{note4}). Shaded areas show energies at which primary electrons ($e^{\uparrow\downarrow}$) and holes ($h^{\uparrow\downarrow}$) are excited by the 1.5 eV pump \cite{Melnikov_PRL11}. (b) Experimental scheme and coordinate frame: $\mathbf{\hat{x}}=[100]$ and $\mathbf{\hat{y}}=[010]$ are equivalent easy axes in Fe(001).}
\end{figure}

The spin-dependent Seebeck effect at a FM/NM interface is governed by (i) the difference in the electronic distribution function $\Delta f(\mathcal{E})$ on both sides of the interface, which depends on the electron energy $\mathcal{E}$, combined with (ii) a spin- and energy-dependent interface transmittance $T$. We have obtained $T$ for the Fe/Au(001) interface by \emph{ab initio} quantum transport calculations \cite{note9}. Thick curves in Fig.~\ref{Fig1}a show the momentum-averaged transmission probability $T^{\uparrow \downarrow}_{FA}$ of majority ($\uparrow$) or minority ($\downarrow$) carriers moving from Fe to Au. For thermalized HC %obeying the Fermi-Dirac distribution
with an electron temperature $T_e(z)$ \cite{Choi_Cahill_NatPhys15}, the gradient $\nabla T_e$ is decisive, whereas the corresponding HC reside within $\pm k_BT_e\lesssim100$~meV around the Fermi energy $\mathcal{E}_F$. Therefore, the spin-polarized current generated at the interface is determined by $\partial T_{FA}^{\uparrow\downarrow}/\partial\mathcal{E}$ at $\mathcal{E}_F$.

Both the magnitude and spin polarization of this current can be enhanced by employing non-thermal HC distributions. The 1.5~eV laser pulses excite primary HC in Fe at energies indicated by shaded areas (Fig.~\ref{Fig1}a) \cite{Melnikov_PRL11}. After few femtoseconds \cite{Zhukov_PRL04,Kaltenborn_PRB14}, the \emph{e-e} scattering leads to the generation of secondary HC which form a non-thermal distribution $f(\mathcal{E})$ \cite{Lisowski_APA04} within $|\mathcal{E}-\mathcal{E}_F|\lesssim$1.5~eV. Among these HC, only majority electrons $e^\uparrow$ with $\mathcal{E}\gtrsim \mathcal{E}_F+$0.3~eV have large transmittance $T^\uparrow_{FA}$ (Fig.~\ref{Fig1}a). This generates a flux of $e^\uparrow$ from Fe to Au forming a pulse of charge ($\mathbf{j}$) and spin ($\mathbf{j^S}$) current in Au. Components $j^S_{z,i}, i=x,y$ define the orientation of the SC spin component while the average electron velocity $\mathbf{v^e}$ is along the normal to interface $\mathbf{\hat{z}}$ (Fig.~\ref{Fig1}b). The collective displacement of "cold" non-spin polarized electrons near $\mathcal{E}_F$ in Au can screen the charge but not spin component of this HC current. Remarkably, the spin-dependent interface transmittance facilitates the demagnetization of a FM layer by superdiffusive spin transport \cite{battiato_PRL10,battiato_PRb12}. Further, since \emph{non-thermalized HC} are essential for NSS, the pulse duration will be primarily determined by the HC thermalization time in FM $\tau^{FM}_{ee}\lesssim0.5$~ps \cite{Rhie_PRL03,Mueller_PRB13}. Thus, on the basis of our calculations, we expect the generation of ultrashort and intense SC pulses, which is demonstrated below.

\begin{figure} \centering
 \includegraphics[width=1\columnwidth]{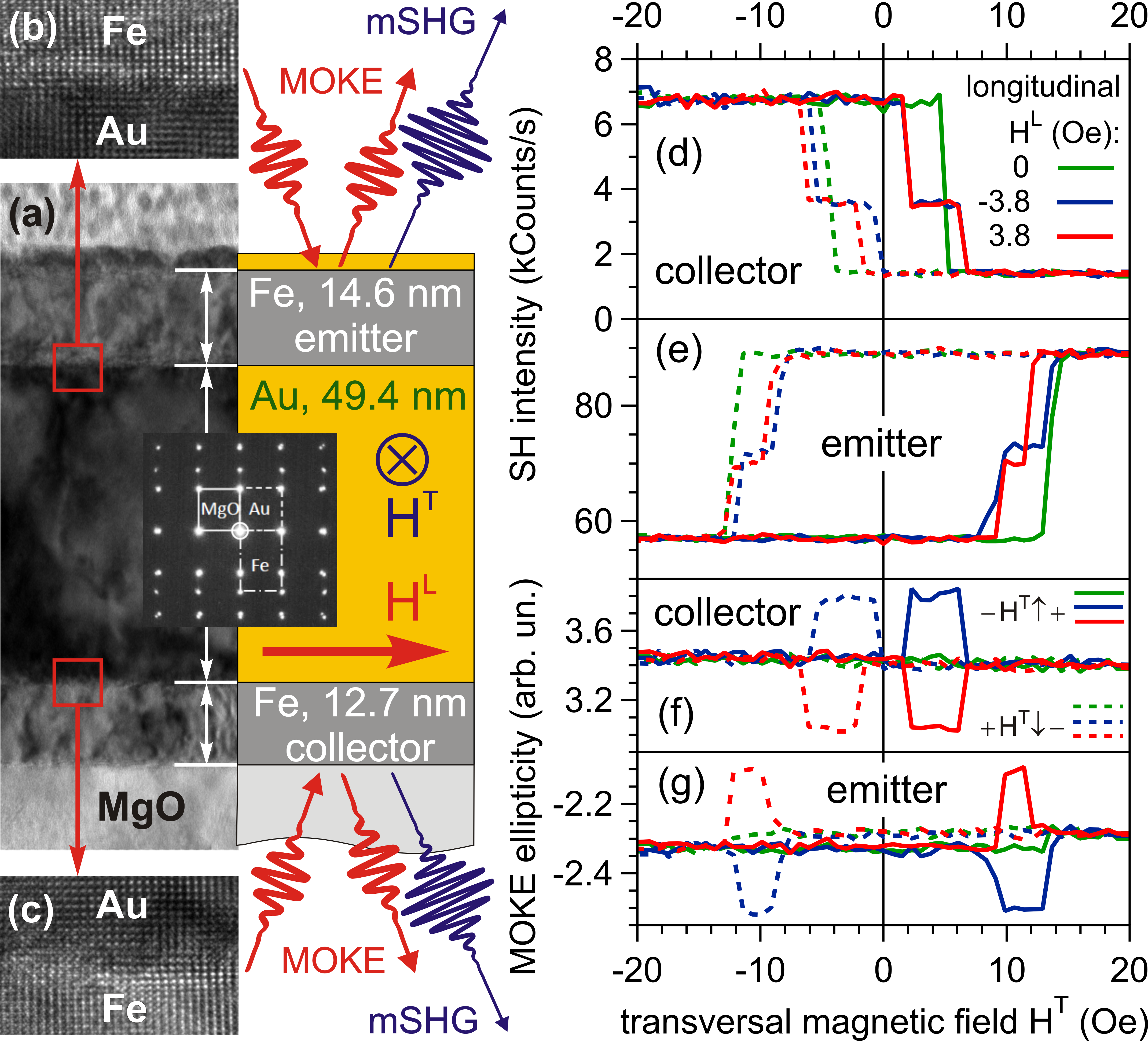}
\caption{\label{Fig2} Cross-section STEM image of Fe/Au/Fe/MgO(001) stack capped with 3~nm of Au (a) obtained using TITAN 80-300 (FEI, USA) combined with the experimental scheme. Electron diffraction (inset) reveals epitaxial growth of layers. HRTEM images of flat Fe/Au (b) and Au/Fe (c) interfaces. Hysteresis loops measured by mSHG (d, e) in \emph{p}-in, \emph{p}-out polarization geometry and by p-in MOKE (f, g) in the bottom (collector) (d, f) and top (emitter) (e, g) Fe layers vs. the transversal external magnetic field $H^T\parallel\mathbf{\hat{y}}$ for the indicated values of the longitudinal field $H^L\parallel\mathbf{\hat{x}}$, as shown in (a).}
\end{figure}

To achieve steady control of propagation direction and spin polarization of SC pulses, we combine the pseudo-spin valve concept with the pump-probe approach developed in Ref.~\cite{Melnikov_PRL11}, see Fig.~\ref{Fig1}b. The two Fe layers in the epitaxial Fe/Au/Fe/MgO(001) stack (Fig.~\ref{Fig2}a) are magnetically decoupled and can be optically excited/probed independently owing to the 50 nm-thick Au spacer \cite{note6}. The samples were fabricated following Ref.~\cite{Melnikov_PRL11} and examined with the scanning transmission electron microscopy (STEM). The excellent epitaxial quality and flatness of interfaces, demonstrated in Fig.~\ref{Fig2} a-c, are essential for comparison with the \emph{ab initio} calculations performed for an atomically sharp interface.

\begin{figure} \centering
 \includegraphics[width=1\columnwidth]{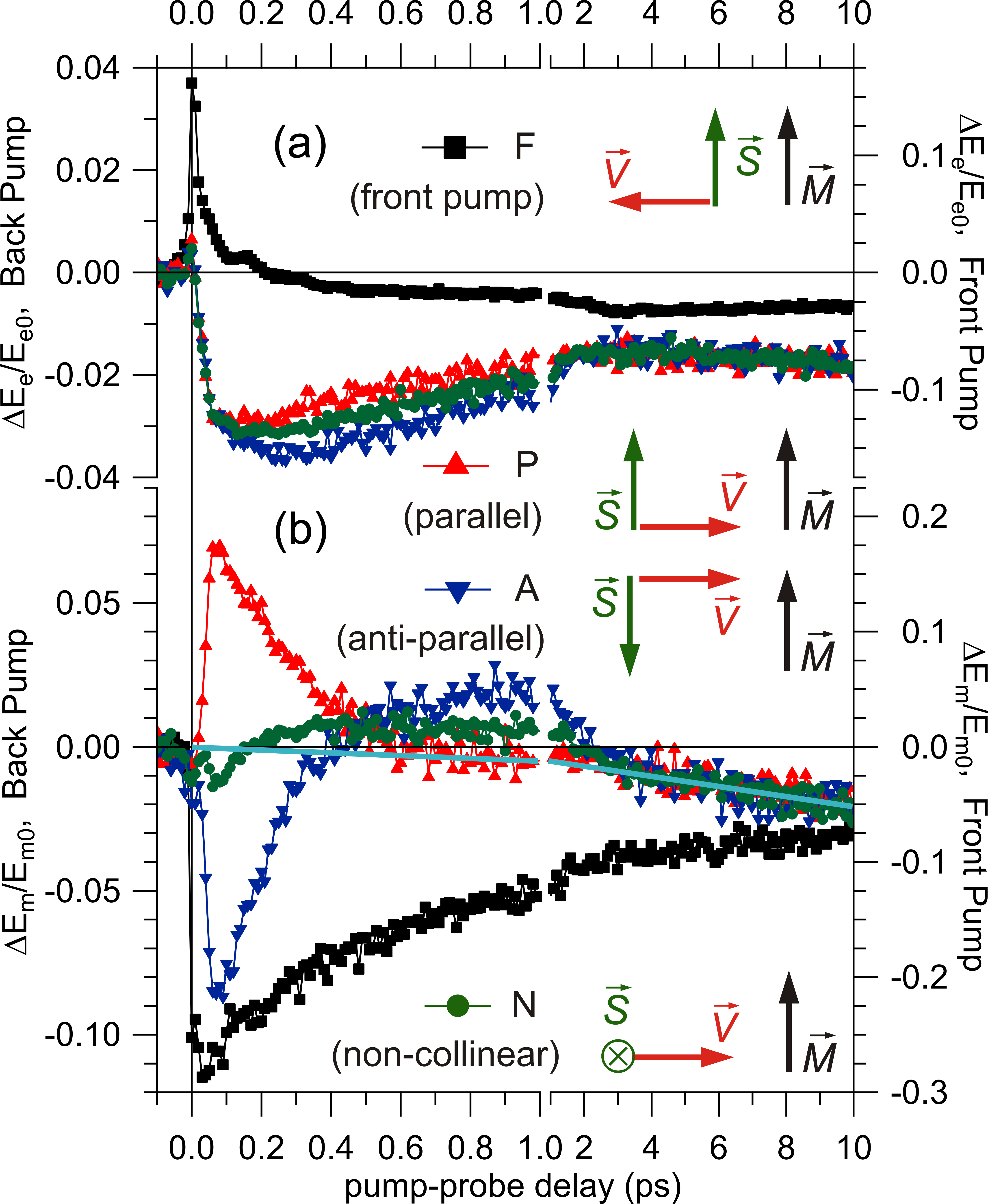}
\caption{\label{Fig3} Variations of non-magnetic and current-induced (a) and magnetization- and SC-induced (b) components of the SH electric field observed in \emph{p-in, p-out} polarization geometry. SC vs. \textbf{M} configurations for the front (F) and back (P, A, N) pumping are shown in the legend (cf. Fig.~\ref{Fig1}b). The solid line represents the common background due to heating-induced demagnetization.}
\end{figure}

\begin{figure} \centering
 \includegraphics[width=1\columnwidth]{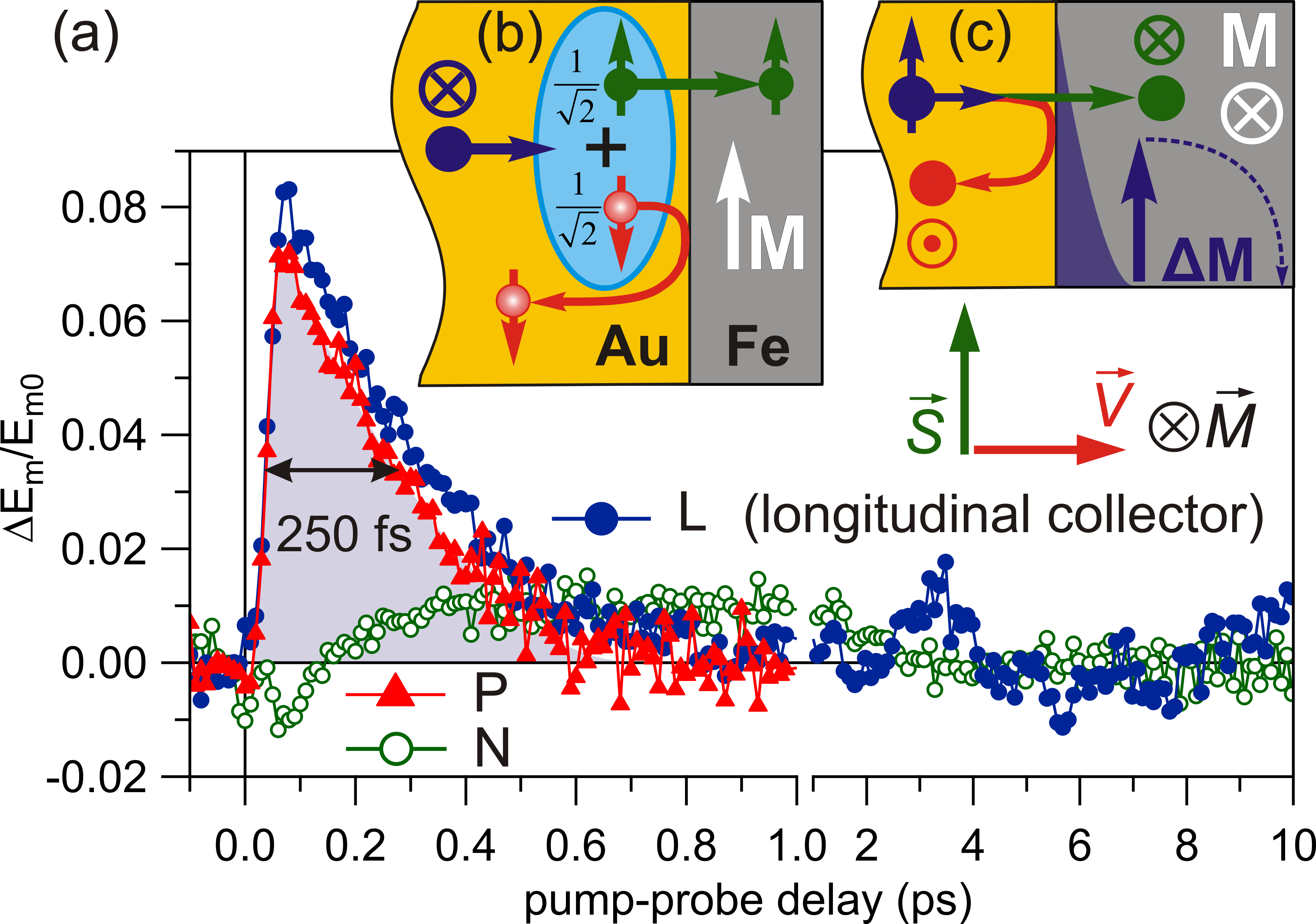}
\caption{\label{Fig4} (a) $\Delta E_m^P$ and $\Delta E_m^N$ with the linear background removed represent the SC pulse shape and the contribution of reflected/transmitted hot electrons, respectively. $\Delta E_m^L$ differs from $\Delta E_m^P$ due to the accumulation of the transversal magnetic moment $\Delta M^C_T$ accumulated in Fe behind the Au/Fe interface [cf. (c) and Eq. (\ref{eq:dEem})]. Hot electrons behavior at the Au/Fe interface is illustrated for N (b) and L (c) SC vs. \textbf{M} configurations (cf. Fig.~\ref{Fig1}). }
\end{figure}

Optical experiments were performed in (either front or back) pump-probe schemes shown in Fig.~\ref{Fig1}b. The 800~nm, 14~fs, 1~MHz output of a cavity-dumped Ti:sapphire oscillator (Mantis, Coherent) was split at a power ratio 4:1 into pump and probe pulses both focused into $\sim10~\mu$m spot resulting in a pump fluence of $\sim$10~mJ/cm$^2$. The magneto-optical Kerr effect (MOKE) was measured with a balanced two-photodiode detector while the SH signal was registered by a photomultiplier \cite{Melnikov_PRL11}. The SH field $\mathbf{E_{2\omega}}=\mathbf{E_e}+\mathbf{E_m}$ consists of "electronic" $\mathbf{E_e}$ and magneto-induced $\mathbf{E_m}$ terms which are, respectively, independent of and proportional to the magnetization $\mathbf{M}$ \cite{pan_PRB89}. In this work (\emph{x,z}) is the plane of incidence and we use \emph{p}-in, \emph{p}-out polarization geometry. Then, $\mathbf{E_m}$ is proportional to $M_y$, the \emph{transversal} magnetization component \cite{pan_PRB89,Melnikov_PRL11} while the MOKE ellipticity is sensitive to the \emph{longitudinal} $M_x$ \cite{Zvezdin1997}. In order to set the magnetization of collector $\mathbf{M^C}$ and emitter $\mathbf{M^E}$ (Fig.~\ref{Fig1}b), transversal ($\mathbf{H^T||\hat{y}}$) and longitudinal ($\mathbf{H^L||\hat{x}}$) magnetic fields were applied. Fig.~\ref{Fig2}d-g shows the hysteresis loops measured vs. $H^T$. For $H^L=0$, both $\mathbf{M^C}$ and $\mathbf{M^E}$ switch abruptly, although at different $H^T$. For $H^L\neq0$, they rotate in two $90^0$ steps, which results in plateaus (Fig.~\ref{Fig2}d,e) corresponding to $\mathbf{M^{C,E}}\uparrow\uparrow\mathbf{H^L}$. At the same field $H^T$, the MOKE ellipticity appears with its sign determined by the direction of $\mathbf{H^L}$ (Fig.~\ref{Fig2}f,g). The step width determined by $H^L$ is set such that we can realize all 16 possible combinations of in-plane $\mathbf{M^C}$ and $\mathbf{M^E}$ parallel to the easy axes. In the pump-probe experiments, the SH intensity was measured in transversal and longitudinal geometries for positive and negative $M_y$ and $M_x$, respectively. From these data, we retrieved $E_e$, $E_m$ for each pump-probe delay $t$ and $E_{e0}$, $E_{m0}$ in the absence of pump \cite{Razdolski_PRB13}. In the following, we discuss relative variations of the SH fields $\Delta E_e/E_{e0}$ and $\Delta E_m/E_{m0}$ which are shown in Figs.~\ref{Fig3}, \ref{Fig4}.

In equilibrium, $\mathbf{E_e}=\mathbf{E_e^{int}}$ and $\mathbf{E_m}=\mathbf{E_m^{int}}$ originate from the \emph{interface} dipole polarization \cite{note2} $P^{2\omega}_i=\chi^{(2)}_{ijk}E^\omega_jE^\omega_k+
\chi^{(2,m)}_{ijk,y}E^\omega_jE^\omega_kM^C_y$, where $i,j,k=x,z$. However, after fs-laser excitation, the presence of a HC current \emph{breaks the inversion symmetry}. This leads to the appearance of \emph{bulk} dipole current- and SC-induced polarizations, $\chi^{(2,C)}_{ijkz}E^\omega_jE^\omega_kj_z$ and $\chi^{(2,SC)}_{ijkz,y}E^\omega_jE^\omega_kj^S_{z,y}$ \cite{note5} contributing to $\mathbf{E_{2\omega}}$ as:
\begin{equation}
    \label{eq:EmSC}
\mathbf{E_e^{C}}\propto j_z\propto v_z,~~~~~~~~\mathbf{E_m^{SC}}\propto j^S_{z,y}\propto v_z s_y,
\end{equation}
where $j_{z,y}^S$ is the SC component with the electron velocity $\mathbf{v}$ along \emph{z} and spin $\mathbf{s}$ along the \emph{y} axis  \cite{note11}. Thus, in addition to the modulation of the interface contributions, the total pump-induced SH variations $\Delta\mathbf{E_{e,m}}$ contain bulk terms (\ref{eq:EmSC}) which provide direct access to the current and SC:
\begin{equation}
    \label{eq:dEem}
\Delta\mathbf{E_{e}}=\Delta\mathbf{E_{e}^{int}}+\mathbf{E_{e}^{C}},~~~~
\Delta\mathbf{E_{m}}=\Delta\mathbf{E_{m}^{int}}+\mathbf{E_{m}^{SC}}.
\end{equation}

This is demonstrated in Fig.~\ref{Fig3} where we induce the HC current from the front (F) or back side of the sample. In the latter case, we consider parallel (P), anti-parallel (A), and orthogonal (N) alignment of $\mathbf{M^E}$ with respect to the transversal $\mathbf{M^C}$. First, we discuss the pronounced characteristic features at $50<t<400$~fs $\sim\tau^{FM}_{ee}$. In the F configuration, we observe $\Delta E_e^F>0$ and $\Delta E_m^F<0$. Then we reverse $\mathbf{v}$ keeping the direction of $\mathbf{s}$ (determined by $\mathbf{M^E}$) and find in this P configuration $\Delta E_e^P<0$ and $\Delta E_m^P>0$, in agreement with the sign change of $j_z$ and $j_{z,y}^S$ in Eq.~(\ref{eq:EmSC}). Then, we again reverse $j_{z,y}^S$ (but not $j_z$) now keeping $v_z$ but changing the sign of $s_y$ (A configuration). This only slightly affects $\Delta E_e$ (Fig.~\ref{Fig3}a) but $\Delta E_m$ changes its sign: $\Delta E_m^A<0$ (Fig.~\ref{Fig3}b). Lastly, we rotate $\mathbf{s}$ by 90$^0$ (N configuration) to set $j_{z,y}^S=0$, $E_m^{SC}=0$ [cf. Eq.~(\ref{eq:EmSC})] and obtain no sizable $\Delta E_m^N$. Thus, we unambiguously observe spin currents owing to the dominant role of $\mathbf{E_m^{SC}}$ in the mSHG response at short delays.

At $t\gg\tau^{FM}_{ee}$, the SC pulse is over and we expect a slow diffusion of thermalized electrons heating up the collector and thus reducing $\Delta\mathbf{E_{m}^{int}}\propto\Delta M^C_y$. This agrees well with the observed $\Delta E_m(t)$ coinciding at $2<t<10$~ps (Fig.~\ref{Fig3}b). Subtracting this common trend taken as $-\alpha t$ (solid line) from $\Delta E_m^P$, we obtain a \emph{unipolar} trace (Fig.~\ref{Fig4}a) with a sharp onset in contrast to the \emph{bipolar} behavior of $\Delta E_m^A$ at $0<t<2$~ps (Fig.~\ref{Fig3}b). To understand the origin of this striking difference and relate the data to the SC pulse shape, it is essential to discuss the interaction of HC with the Au/Fe interface, which is determined by the interface transmittance $T_{AF}$ (Fig.~\ref{Fig1}a, thin curves). Firstly, we will consider "open" (P) and "closed" (A) states of the spin valve widely used in GMR (giant magneto-resistance) devices and identify contributions of transmitted and reflected HC to mSHG. After that we will turn to N and L configurations (Fig.~\ref{Fig4}b,c) with $\mathbf{M^E}\perp\mathbf{M^C}$ having high potential functionality for spin transfer torque applications \cite{Razdolski_XXX}.

Averaging the calculated HC transmittance over the energy of emitted HC we obtain $T^{P}=\langle T^{\uparrow}_{AF}\rangle\approx0.95$ and $T^{A}=\langle T^{\downarrow}_{AF}\rangle\approx0.25$. This means that in P or A configurations the collector works as an effective HC \emph{absorber} or \emph{reflector}. In the latter case, the reflected SC and transient spin density in Au near the interface modify $\mathbf{E_{m}^{SC}}$ \cite{note7} and $\mathbf{E_{m}^{int}}$, respectively. The observed bipolar shape of $\Delta E_m^A(t)$ is similar to that measured in Ref.~\cite{Melnikov_PRL11} at the Au surface where the average HC reflection $R=1$ is close to $R^A=1-T^A\approx0.75$. This similarity points at the negligibility of $\Delta\mathbf{E_{m}^{int}}$ originating from transient reduction of $M^C$ driven by the spin injection which is ruled out in Ref. \cite{Melnikov_PRL11}. In P configuration, the transmitted angular momentum is about four times larger than in A. However, it is distributed within the electron inelastic mean free path in Fe $\lambda_{Fe}^{\uparrow}\gg\lambda_{Fe}^{\downarrow}$ \cite{Zhukov_PRB06} thus providing even smaller $\Delta M^C$ at the interface. Therefore, all contributions to $\Delta E_m^P$ apart from $E_m^{SC}(t)$ are small \cite{note8} and $\Delta E_m^P(t)$ in Fig.~\ref{Fig4}a represents the temporal profile of the SC pulse in the entire range of delays.

For ballistic electrons traveling with the Fermi velocity 1.4~nm/fs \cite{Brorson_PRL87} and random angular distribution the average time of propagation through the Au layer $\tau_b\approx70$~fs. The maximum of $\Delta E_m^P(t)$ observed at this delay thus suggests the ballistic character of HC transport \cite{Melnikov_PRL11} and indicates the ballistic length in Au $\lambda_{Au}\gtrsim100$~nm for electrons with $0.3<\mathcal{E}-\mathcal{E}_F<1.5$~eV. Owing to that, we conclude that the SC pulse shape given by $\Delta E_m^P(t)$ closely reproduces the dynamcs of HC emission with the decay time of about 250 fs (Fig.~\ref{Fig4}a), which is similar to the HC thermalization time $\tau^{Fe}_{ee}\approx200$~fs \cite{Alekhin2016PhD}. Thus, this particular transient structure of $\Delta E_m^P$ is a strong evidence for the NSS mechanism of SC generation considered above.

We now turn to L configuration (Fig.~\ref{Fig4}c) where the longitudinal $\mathbf{M^C}$ ensures the absence of demagnetization effects in $\mathbf{E_{m}^{int}}$ while the transversal $\mathbf{M^E}$ provides a non-zero $\mathbf{E_{m}^{SC}}$. The striking similarity of $\Delta E_m^P$ and $\Delta E_m^L$ at $t<1$~ps (Fig.~\ref{Fig4}a) indicates a negligible contribution of reflected HC spins. We treat $\mathbf{s}\perp\mathbf{M^C}$ representing its wave function $|\Psi_{\perp}\rangle=(|\Psi_{\uparrow}\rangle + |\Psi_{\downarrow}\rangle)/\sqrt{2}$ as a superposition of the spin eigenfunctions \cite{Stiles_PRB02} corresponding to $\mathbf{s}\uparrow\uparrow\mathbf{M^C}$ and $\mathbf{s}\downarrow\uparrow\mathbf{M^C}$, which were analyzed above for P and A configurations. Considering diagonal reflection and transmission operators, we obtain that the transmittance for electrons with $\mathbf{s}\perp\mathbf{M^C}$ is the \emph{average of transmittance} for electrons with $\mathbf{s}\uparrow\uparrow\mathbf{M^C}$ \emph{and} $\mathbf{s}\downarrow\uparrow\mathbf{M^C}$. A limiting case of $T^P=1$ and $T^A=0$ is illustrated in Fig.~\ref{Fig4}b: only the spin-up (-down) component of the incoming wave is transmitted (reflected). In other words, all \emph{orthogonal HC spins} interacting with FM \emph{rotate} to parallel and anti-parallel with equal probabilities. Subsequently, parallel spins are transmitted and anti-parallel reflected, i.e. the two spin components are separated in space like in the \emph{Stern-Gerlach experiment}.

In the general case, not all HC spins rotate at the interface (spin polarization rotates by less than $90^0$). Spin components of reflected and transmitted SC parallel to $\mathbf{M^C}$ ("rotated spins") are given \cite{Stiles_PRB02} by $s^R_{||}=(R^P-R^A)|s_{\perp}|/2$ and $s^T_{||}=(T^P-T^A)|s_{\perp}|/2$, respectively, where $s_{\perp}$ denotes the polarization of incoming SC with $\mathbf{s}\perp\mathbf{M^C}$. The residual perpendicular spin components ("conserved spins") are $|s^R_{\perp}|\leq\sqrt{R^PR^A}|s_{\perp}|$ and $|s^T_{\perp}|\leq\sqrt{T^PT^A}|s_{\perp}|$.
These $s^{R,T}_{||}$ and $s^{R,T}_{\perp}$ are addressed in N ($\mathbf{s}\parallel\mathbf{\hat{x}}$ and $\mathbf{M^C}\parallel\mathbf{\hat{y}}$) and L ($\mathbf{M^C}\parallel\mathbf{\hat{x}}$ and $\mathbf{s}\parallel\mathbf{\hat{y}}$) configurations, respectively. Since all terms in Eq.~(\ref{eq:dEem}) are linear with $T^i$, $R^i$ and all $s_y$-dependent terms change sign when switching from P to A configuration, we arrive at $\Delta E^N(t)=(\Delta E^P+\Delta E^A)/2$. Experimentally this equation holds for both $\Delta E_{e,m}$ within error bars (not shown) in line with Ref.~\cite{Stiles_PRB02}. Thus, the positive $\Delta E_m^N$ observed at $0.5<t<2$~ps is generated by reflected HC spins rotated into the anti-parallel direction. The negative peak observed in $\Delta E_m^N$ at 100~fs characterizes the effect of transmitted HC on $M^C$. The negligible effect of reflected HC spins in L configuration agrees well with the calculated $R^{P,A}$ for which the residual $|s^R_{\perp}|$ does not exceed $|s_{\perp}|/5$: most of these spins are rotated to $\mathbf{s}\downarrow \uparrow \mathbf{M^C}$ and thus do not contribute to $\Delta E^L_m$.

The angular momentum conservation upon interaction with the Au/Fe interface leads to the emergence of $\Delta M^C_{\perp}$ in a thin Fe layer adjacent to the interface. This is illustrated in Fig.~\ref{Fig4}c for L configuration in the limiting case of $T^P=1$ and $T^A=0$.  Although for the calculated $T^{P,A}$ significant $s^T_{\perp}$ up to $s_{\perp}/2$ can survive, the quantum decoherence of the $|\Psi_{\uparrow}\rangle + |\Psi_{\downarrow}\rangle$ superposition results in the angular momentum accommodation within $\lambda_{Fe}^{\downarrow}$. This \emph{ultrafast spatially confined spin transfer torque} effect inducing $\Delta M^C_y$ at the interface is responsible for the small deviation of $\Delta E_m^L$ from $\Delta E_m^P$ observed in Fig.~\ref{Fig4}a at $t<0.5$~ps. Subsequently, it leads to the excitation of several lowest standing spin wave modes in the collector, opening a pathway for studies of spatially non-uniform precessional spin dynamics expanding towards the THz domain. Being the origin of the non-monotonous behavior of $\Delta E_m^L$ observed at the long time scale in Fig.~\ref{Fig4}a, this dynamics is discussed elsewhere \cite{Razdolski_XXX}.

Summarizing, we have demonstrated the generation of ultrashort SC pulses in Fe/Au/Fe epitaxial layers using the nonlinear-optical technique developed for direct monitoring of SC. The measured pulse shape agrees well with the proposed non-thermal spin-dependent Seebeck effect and indicates a nearly-ballistic character of HC transport in Au. The SC pulse duration ($\sim$250 fs) is determined by the HC thermalization time in Fe. We have shown the large difference in transmittance across the Au/Fe interface for the majority and minority electrons with $\mathcal{E}-\mathcal{E}_F<1.5$~eV. This results in a high spin rotation efficiency ($\sim70\%$) at the interface, where the transmitted (reflected) SC loses its orthogonal spin component and becomes polarized parallel (anti-parallel) to the Fe magnetization. These findings facilitate the development of metal-based sources of ultrashort SC pulses and reflective/transmittive spin polarizers/rotators for ultrafast spintronics.

The authors thank T.O. Wehling, A.I.~Lichtenstein, P.M.~Oppeneer, M.~Weinelt, and T.~Kampfrath for fruitful discussions and M.~Wolf for continuous support. Funding by the Deutsche Forschungsgemeinschaft through ME 3570/1 and Sfb 616, the EU 7-th framework program through CRONOS and ACMOL (grants No. 280879 and 618082), and Science Foundation Ireland (grant No. 14/IA/2624) is gratefully acknowledged.


%\bibliography{FeAuFe_SC_2015_02_bib}

\begin{thebibliography}{10}
\bibitem{note4}
The difference between $T^{\uparrow \downarrow}_{FA}$ and $T^{\uparrow
  \downarrow}_{AF}$ is due to different numbers of incoming states in Fe and Au
  to which the number of interface states is normalized.
\bibitem{note9}
We have calculated the full spin-dependent transmission matrix between all
  incoming states in Fe and the outgoing states in Au for the scattering region
  consisting of 12 atomic layers of Fe and same of Au using the non-equilibrium
  Green's function technique implemented in the Smeagol code
  \cite{Soler_JPCM02,Rocha_PRB06} at the level of the local density
  approximation.
\bibitem{note6}
A semi-transparent alignment area has been fabricated for the beam overlapping
  and precise determination of $t=0$ by the SHG cross-correlation technique
  \cite{Melnikov_PRL11}.
\bibitem{note2}
Here we neglect the bulk quadrupole contribution which was shown to be small
  with respect to the surface dipole one in Au \cite{Wang_PRB09}.
\bibitem{note7}
Because of the velocity sign reversal, contributions of incoming and reflected
  SC to $\mathbf{E_{m}^{SC}}$ have opposite signs, according to Eq.~(1).
\bibitem{note8}
The current- and SC-induced contributions from the collector are neglected due
  to much smaller nonlinearity (almost two orders of magnitude smaller SH
  intensity) and inelastic mean free path in Fe \cite{Zhukov_PRB06}, as
  compared to Au.
\end{thebibliography}
\bibliographystyle{prsty}

\end{document}